\PassOptionsToPackage{table}{xcolor}
\documentclass[conference,a4paper]{APSIPA2025}
\usepackage[table]{xcolor}
\usepackage{amsmath}

\usepackage{graphicx}
\usepackage{multirow}
\usepackage{multicol}
\usepackage{threeparttable}
\usepackage{url}

\usepackage{subcaption}

\usepackage{geometry}
\usepackage{fancyhdr}

\fancypagestyle{firststyle}{
  \fancyhf{}
  \fancyhead[C]{2025 Asia Pacific Signal and Information Processing Association Annual Summit and Conference (APSIPA ASC)}
}
\let\svthefootnote\thefootnote
\newcommand\blankfootnote[1]{%
  \let\thefootnote\relax\footnotetext{#1}%
  \let\thefootnote\svthefootnote%
}

\begin{document}

\title{Robust Localization of Partially Fake Speech: Metrics and Out-of-Domain Evaluation}

\author{
\authorblockN{
Hieu-Thi Luong\authorrefmark{1},
Inbal Rimon\authorrefmark{2},
Haim Permuter\authorrefmark{2}, 
Kong Aik Lee\authorrefmark{3}
and Eng Siong Chng\authorrefmark{1}
}

\authorblockA{
\authorrefmark{1}
Nanyang Technological University, Singapore \\
hieuthi.luong@ntu.edu.sg, aseschng@ntu.edu.sg}

\authorblockA{
\authorrefmark{2}
Ben Gurion University, Be’er Sheva, Israel \\
inbalri@post.bgu.ac.il, haimp@bgu.ac.il}

\authorblockA{
\authorrefmark{3}
The Hong Kong Polytechnic University, Hong Kong SAR, China \\
kong-aik.lee@polyu.edu.hk}}

\maketitle
\thispagestyle{firststyle}
\pagestyle{fancy}

\begin{abstract}
Partial audio deepfake localization poses unique challenges and remain underexplored compared to full-utterance spoofing detection. While recent methods report strong in-domain performance, their real-world utility remains unclear. In this analysis, we critically examine the limitations of current evaluation practices, particularly the widespread use of Equal Error Rate (EER), which often obscures generalization and deployment readiness.
We propose reframing the localization task as a sequential anomaly detection problem and advocate for the use of threshold-dependent metrics such as accuracy, precision, recall, and F1-score, which better reflect real-world behavior.
Specifically, we analyze the performance of the open-source Coarse-to-Fine Proposal Refinement Framework (CFPRF), which achieves a 20-ms EER of 7.61\% on the in-domain PartialSpoof evaluation set, but 43.25\% and 27.59\% on the LlamaPartialSpoof and Half-Truth out-of-domain test sets. Interestingly, our reproduced version of the same model performs worse on in-domain data (9.84\%) but better on the out-of-domain sets (41.72\% and 14.98\%, respectively).
This highlights the risks of over-optimizing for in-domain EER, which can lead to models that perform poorly in real-world scenarios.
It also suggests that while deep learning models can be effective on in-domain data, they generalize poorly to out-of-domain scenarios, failing to detect novel synthetic samples and misclassifying unfamiliar bona fide audio.
Finally, we observe that adding more bona fide or fully synthetic utterances to the training data often degrades performance, whereas adding partially fake utterances improves it.
\end{abstract}

\section{Introduction}
\blankfootnote{
\textsuperscript{\dag}Dr. Hieu-Thi Luong is funded by RIE2025 NRF International Partnership Funding Initiative. This research is supported by the National Research Foundation, Singapore, under the AI Singapore Programme (AISG Award No.: AISG2-TC-2023-011-SGIL). Any opinions, findings and conclusions or recommendations expressed in this material are those of the author(s) and do not reflect the views of National Research Foundation, Singapore.
}

While full-utterance spoofing has been widely studied \cite{wu2017asvspoof, muller2024mlaad,wang2025asvspoof}, partial audio deepfakes, where only segments of an utterance are manipulated, are more challenging due to their subtlety \cite{zhang2022partialspoof,yi2023add}. These localized forgeries can evade traditional spoofing detection systems that rely on global audio characteristics, making them particularly threatening in scenarios such as voice-controlled authentication and disinformation generation \cite{luong2025llamapartialspoof}.
The performance evaluation in partial deepfake scenarios remains inconsistent across the literature \cite{yi2023add,cai2024av}. Zhang et al. \cite{zhang2022partialspoof} used segment-based Equal Error Rate (EER) as the primary metric, while Cai et al. \cite{cai2024av} adopted Average Recall (AR) and Average Precision (AP). Although these metrics are threshold-independent, which simplifies evaluation, they can be difficult to interpret in the context of real-world deployment.
Current localization models report outstanding results on in-domain datasets \cite{luong2025llamapartialspoof,wu2024coarse}. For example, Cai et al. \cite{cai2024integrating} achieved a 20-ms segment-based EER of 0.064\% on the ADD 2023 Challenge Track 2 development set \cite{yi2023add}, and 1.74\% on the PartialSpoof evaluation set \cite{zhang2022partialspoof}. However, cross-domain evaluations reveal a significant drop in performance, 15.2\% EER on PartialSpoof and 45.34\% on ADD, indicating that the models struggle to generalize beyond their training domain \cite{muller22_interspeech}.
These results raise concerns about whether these systems are sufficiently robust and reliable for real-world use.

To create a shift in research towards more robust models \cite{yang2024robust,zhu2024slim,doan2024trident}, we recently released the LlamaPartialSpoof dataset \cite{luong2025llamapartialspoof}, which contains high-quality fully and partially fake speech designed for out-of-domain evaluation.
In this work, we discuss the shortcomings of the current evaluation methodology \cite{pascu2024easy}, and propose a simpler framing that better assesses system robustness that helps bridge the gap between research and deployment.
The contributions of this paper are as follows: (1) Metric Redefinition: We argue that the metrics used in recent research are problematic and can be misleading, particularly on out-of-domain data. To address this, we propose reframing the localization task as a standard anomaly detection problem and advocate the use of conventional metrics such as accuracy, precision, recall, and F1; (2) Generalization Limitations of SSL Models: We show that current self-supervised learning (SSL)-based models fail to generalize to out-of-domain data. These models struggle not only to detect novel synthetic speech but also to misclassify out-of-domain bona fide audio; (3) Impact of Training Data: Finally, we show that increasing the volume of training data does not necessarily improve performance and, in some cases, may even degrade it.

\section{Partially Fake Speech Localization}

\begin{table}[t]
    \caption{Partially Fake Speech Datasets for In-Domain and Out-of-Domain Training and Evaluation}
    \label{tab:data}
    \centering    
    \scalebox{0.95}{
    \begin{tabular}{l|l|rrrr}
         \hline \hline
         \multirow{2}{*}{Dataset} & \multirow{2}{*}{Language} & \multicolumn{3}{c}{Number of Utterance}  \\ 
          & & Bonafide & Full Fake  & Partial Fake \\ \hline
         PartialSpoof & English \\ 
         - train & & 2580 & 0 & 22,800 \\
         - dev & & 2548 & 0 & 22,296 \\
         - eval & & 7,355 & 0 & 63,882 \\ \hline
         LlamaPartialSpoof & English & 10,573 & 33,461 & 32,194 \\
         - (subset) train & & 2,391 & 13,991 & 13,498 \\
         - (subset) test & & 8,182 & 19,470 & 18,696 \\ \hline
         Half-Truth & Chinese &\\
         - test & & 0 & 0 & 9072 \\ \hline

    \end{tabular}} \\ 
    \vspace{-3mm}
    \label{tab:datasets}
\end{table}

\subsection{Problem Definition}

Given an input audio of $N$ frames, each of duration $d$ seconds, the task of partially fake localization aims to produce an output sequence $\tilde{Y}=\{\tilde{y}_1, \tilde{y}_2, ..., \tilde{y}_N\}$ that best matches the ground truth labels $Y=\{y_1, y_2, ..., y_N\}$. Each frame label can be either 0, denoting a bona fide segment, or 1, denoting a fake segment.
Note that this labeling scheme is the inverse of that used in several previous studies \cite{zhang2022partialspoof,wu2024coarse}. We adopt this formulation to better align the task with standard practices in anomaly detection \cite{coletta2025anomaly} and to reduce the risk of misinterpreting the evaluation results.
In this work, we label segments that contain any amount of fake audio as fake, instead of requiring a majority.

\subsection{Datasets and Generalization}
Compared to audio deepfake detection, the task of localizing small fake segments within longer utterances is less explored, partly due to the scarcity of publicly available datasets. This is largely because creating such datasets requires greater expertise.
We trained our systems on the train set of PartialSpoof \cite{zhang2022partialspoof} and used its eval set for in-domain evaluation.
The main focus of this work is to investigate whether localization systems can generalize to unseen scenarios, which is a key indicator of their readiness for real-world deployment.
To assess out-of-domain generalization, we additionally evaluated our models on two additional datasets: LlamaPartialSpoof \cite{luong2025llamapartialspoof} and Half-Truth \cite{Yi2021halftruth}.
Table \ref{tab:datasets} shows the number of utterances included in each dataset. Notably, the LlamaPartialSpoof dataset contains bona fide samples, as well as both fully fake and partially fake utterances, whereas the test set of the Half-Truth dataset includes only partially fake utterances.

\subsection{Fake Speech Localization Models}
In this paper, we use two SSL-based fake speech localization systems to examine various aspects of evaluating and improving robustness in out-of-domain scenarios.
Specifically, we utilize the Multi-resolution Model (MRM)\footnote{\url{https://github.com/hieuthi/MultiResoModel-Simple}}, introduced alongside the PartialSpoof dataset \cite{zhang2022partialspoof}, and the Coarse-to-Fine Proposal Refinement Framework (CFPRF)\footnote{\url{https://github.com/ItzJuny/CFPRF}} \cite{wu2024coarse}.
MRM processes input at multiple temporal resolutions, ranging from 20 ms to 640 ms, using a hierarchical architecture that enables the model to learn temporal patterns at varying scales.
CFPRF employs a two-stage architecture: a 20-ms Frame-level Detection Network (FDN) first identifies coarse candidate regions based on inconsistency cues between real and fake frames, followed by a Proposal Refinement Network (PRN) that fine-tunes boundary predictions. CFPRF has reported state-of-the-art results on in-domain localization tasks. These models were trained using the default setting of their respectable setups.

\begin{table}[t]
    \caption{Evaluation results of the CFPRF model using genuine and fake scores calculated on the PartialSpoof evaluation set at the 20-ms resolution.}
    \centering
    \label{tab:cfprf}

    \scalebox{1.0}{
    \begin{tabular}{l|r|rrrr}
         \hline \hline
         \multirow{2}{*}{Score} & \multirow{2}{*}{EER} & \multicolumn{4}{c}{Threshold at EER} \\
          &  & Acc. & Pre.\cellcolor{gray!25} & Rec.\cellcolor{gray!25} & F1\cellcolor{gray!25} \\ \hline
         genuine score (reported) & 7.41 & 92.58 & 95.23\cellcolor{gray!25} & 92.59\cellcolor{gray!25} & 93.89\cellcolor{gray!25} \\
         $1- \text{genuine score}$ (ours) & 7.61 & 92.39 & 88.77\cellcolor{gray!25} & 92.39\cellcolor{gray!25} & 90.54\cellcolor{gray!25} \\ \hline
    \end{tabular}}
    \vspace{-3mm}
\end{table}

\subsection{Evaluation Metrics}

Current approaches to deepfake audio detection typically follow the anti-spoofing evaluation methodology, relying on metrics such as EER or tandem Detection Cost Function (t-DCF) \cite{kinnunen2020tandem} to quickly assess model performance. However, these metrics postpone the selection of a decision threshold and reduce the interpretability of results, making it more difficult to evaluate system performance in practical scenarios.
While closely related, fake speech detection is a broader and simpler task compared to speaker verification spoofing, as it does not necessarily involve the imitation of a specific target speaker. In the case of partially fake speech, the problem can be naturally framed as a sequential anomaly detection task. This allows for the use of straightforward and interpretable metrics such as accuracy, precision, recall, and F1-score.

Table \ref{tab:cfprf} shows the evaluation results of the CFPRF model released by the authors \cite{wu2024coarse}. The first row reports results based on the model’s output score (interpreted as a genuine score), while the second row uses $1 - \text{score}$ (interpreted as a fake score). In these two scenarios,  the meaning of EER and accuracy remains the same, while precision, recall, and F1 differ depending on which class is defined as the positive class. This accounts for the significant differences observed in the precision results, beyond small mismatches that may arise from differences in segment labeling.
This highlights how the current evaluation framing can misrepresent model performance and lead to confusion. For example, readers might expect the fake speech localization system to achieve 95.23\% precision, when its actual precision is only 88.77\%.

Zhang et al. \cite{zhang23v_interspeech} advocated for the use of Range-based EER, which is essentially a segment-based EER computed over very short segments. This metric enables finer-grained  and fairer comparisons between different setups. However, in this work, we focus on the robustness of the model rather than its finer-grained precision, and therefore leave such evaluations for future work.
Throughout the rest of this paper, all evaluation results are calculated with the fake segment defined as the positive class, in alignment with standard anomaly detection practices.
To support standardized evaluation, we have open-sourced our library of partial spoofing metrics\footnote{https://github.com/hieuthi/partialspoof-metrics}.

\begin{table}[t]
    \caption{EERs at different resolutions.}
    \centering

    (a) PartialSpoof (in-domain) \\
    \scalebox{0.95}{
    \begin{tabular}{l|rrrrrrr}
         \hline \hline
         \multirow{2}{*}{Model} & \multicolumn{7}{c}{Evaluation resolution (s)}  \\
          & 0.02 & 0.04 & 0.08 & 0.16 & 0.32 & 0.64 & Utt. \\ \hline
         MRM & 13.72 & 14.46 & 15.29 & 11.60 & 9.63 & 7.24 & \textbf{1.48} \\
         CFPRF & \textbf{7.61} & \textbf{7.36}\cellcolor{gray!25} & \textbf{6.84}\cellcolor{gray!25} & \textbf{6.04}\cellcolor{gray!25} & \textbf{5.24}\cellcolor{gray!25} & \textbf{4.80}\cellcolor{gray!25} & 1.72\cellcolor{gray!25} \\ 
         reCFPRF & 9.84 & 9.47\cellcolor{gray!25} & 8.73\cellcolor{gray!25} & 7.42\cellcolor{gray!25} & 5.89\cellcolor{gray!25} & 4.88\cellcolor{gray!25} & 1.65\cellcolor{gray!25} \\ \hline
    \end{tabular}}
    \vspace{3mm}
    
    (b) LlamaPartialSpoof (out-of-domain) \\
    \scalebox{0.95}{
    \begin{tabular}{l|rrrrrrr}
         \hline \hline
         \multirow{2}{*}{Model} & \multicolumn{7}{c}{Evaluation resolution (s)}  \\
          & 0.02 & 0.04 & 0.08 & 0.16 & 0.32 & 0.64 & Utt. \\ \hline
         MRM & 46.29 & 45.93 & 45.13 & 43.43 & 40.79 & 37.04 & \textbf{24.50} \\
         CFPRF & 43.25 & 42.85\cellcolor{gray!25} & 42.15\cellcolor{gray!25} & 40.95\cellcolor{gray!25} & 38.95\cellcolor{gray!25} & 36.04\cellcolor{gray!25} & 31.50\cellcolor{gray!25} \\ 
         reCFPRF & \textbf{41.72} & \textbf{41.37}\cellcolor{gray!25} & \textbf{40.77}\cellcolor{gray!25} & \textbf{39.77}\cellcolor{gray!25} & \textbf{37.78}\cellcolor{gray!25} & \textbf{34.10}\cellcolor{gray!25} & 24.81\cellcolor{gray!25} \\ \hline

    \end{tabular}} \\
    \vspace{3mm}
 
    (c) Half-Truth (out-of-domain) \\
    \scalebox{0.95}{
    \begin{tabular}{l|rrrrrrr}
         \hline \hline
         \multirow{2}{*}{Model} & \multicolumn{7}{c}{Evaluation resolution (s)}  \\
          & 0.02 & 0.04 & 0.08 & 0.16 & 0.32 & 0.64 & Utt. \\ \hline
         MRM &  46.48 & 46.29 & 45.50 & 43.39 & 41.83 & 42.93 & -  \\
         CFPRF & 27.59 & 27.98\cellcolor{gray!25} & 28.41\cellcolor{gray!25} & 29.21\cellcolor{gray!25} & 31.03\cellcolor{gray!25} & 33.31\cellcolor{gray!25} & -\cellcolor{gray!25}  \\ 
         reCFPRF & \textbf{14.98} & \textbf{14.86}\cellcolor{gray!25} & \textbf{14.65}\cellcolor{gray!25} & \textbf{14.62}\cellcolor{gray!25} & \textbf{14.97}\cellcolor{gray!25} & \textbf{16.18}\cellcolor{gray!25} & -\cellcolor{gray!25}  \\ \hline

    \end{tabular}} \\
    \vspace{-3mm}
    \label{tab:eer}
\end{table}

\begin{table}[t!]
    \caption{Evaluation results at different resolutions at a specific threshold and recall values.}
    \centering

    (a) PartialSpoof (in-domain) \\    
    \scalebox{0.92}{
    \begin{tabular}{ll|rrr|rrr}
         \hline \hline
         \multirow{2}{*}{Model} & \multirow{2}{*}{Reso.} &  \multicolumn{3}{c|}{Threshold=0.5000} & \multicolumn{3}{c}{Recall=95\%} \\
          & & Pre. & Rec. & F1 & Pre. & Thres. & F1 \\ \hline
          \multirow{7}{*}{MRM} & 0.02 s & 96.73 & 70.28 & 81.41 & 53.95 & 0.0048 & 68.86 \\ 
          & 0.04 s & 97.21 & 71.36 & 82.30 & 51.65 & 0.0024 & 66.97 \\ 
          & 0.08 s & 97.25 & 74.57 & 84.41 & 46.15 & -0.0232 & 62.13 \\ 
          & 0.16 s & 97.57 & 79.05 & 87.34 & 62.55 & 0.0000 & 75.55 \\ 
          & 0.32 s & 98.06 & 84.09 & 90.54 & 80.53 & 0.0116 & 87.19 \\ 
          & 0.64 s & 98.74 & 89.23 & 93.75 & 90.85 & 0.0172 & 92.89 \\ \cline{2-8}
          & Utt. & 99.19 & 99.52 & 99.35 & 99.96 & 0.9264 & 97.43 \\\hline

          \multirow{7}{*}{CFPRF} & 0.02 s & 94.93 & 86.90 & 90.74 & 81.35 & 0.0212 & 87.65 \\ 
          & 0.04 s & 95.07\cellcolor{gray!25} & 87.76\cellcolor{gray!25} & 91.27\cellcolor{gray!25} & 83.66\cellcolor{gray!25} & 0.0304\cellcolor{gray!25} & 88.97\cellcolor{gray!25} \\
          & 0.08 s & 95.29\cellcolor{gray!25} & 89.37\cellcolor{gray!25} & 92.24\cellcolor{gray!25} & 87.44\cellcolor{gray!25} & 0.0556\cellcolor{gray!25} & 91.07\cellcolor{gray!25} \\ 
          & 0.16 s & 95.75\cellcolor{gray!25} & 91.76\cellcolor{gray!25} & 93.71\cellcolor{gray!25} & 92.30\cellcolor{gray!25} & 0.1536\cellcolor{gray!25} & 93.63\cellcolor{gray!25} \\ 
          & 0.32 s & 96.58\cellcolor{gray!25} & 94.29\cellcolor{gray!25} & 95.42\cellcolor{gray!25} & 96.14\cellcolor{gray!25} & 0.4148\cellcolor{gray!25} & 95.57\cellcolor{gray!25} \\ 
          & 0.64 s & 97.78\cellcolor{gray!25} & 96.33\cellcolor{gray!25} & 97.05\cellcolor{gray!25} & 98.18\cellcolor{gray!25} & 0.6396\cellcolor{gray!25} & 96.57\cellcolor{gray!25} \\ \cline{2-8} 
          & Utt. & 99.45\cellcolor{gray!25} & 99.42\cellcolor{gray!25} & 99.44\cellcolor{gray!25} & 99.89\cellcolor{gray!25} & 0.9488\cellcolor{gray!25} & 97.40\cellcolor{gray!25} \\\hline

          \multirow{7}{*}{reCFPRF} & 0.02 s & 96.43 & 81.60 & 88.40 & 64.28 & -0.0192 & 76.69 \\ 
          & 0.04 s & 96.51\cellcolor{gray!25} & 82.80\cellcolor{gray!25} & 89.13\cellcolor{gray!25} & 67.82\cellcolor{gray!25} & -0.0120\cellcolor{gray!25} & 79.15\cellcolor{gray!25} \\
          & 0.08 s & 96.60\cellcolor{gray!25} & 84.92\cellcolor{gray!25} & 90.38\cellcolor{gray!25} & 74.53\cellcolor{gray!25} & 0.0004\cellcolor{gray!25} & 83.53\cellcolor{gray!25} \\ 
          & 0.16 s & 96.86\cellcolor{gray!25} & 88.05\cellcolor{gray!25} & 92.24\cellcolor{gray!25} & 85.84\cellcolor{gray!25} & 0.0396\cellcolor{gray!25} & 90.19\cellcolor{gray!25} \\ 
          & 0.32 s & 97.38\cellcolor{gray!25} & 91.60\cellcolor{gray!25} & 94.40\cellcolor{gray!25} & 94.69\cellcolor{gray!25} & 0.2340\cellcolor{gray!25} & 94.85\cellcolor{gray!25} \\ 
          & 0.64 s & 98.24\cellcolor{gray!25} & 94.71\cellcolor{gray!25} & 96.45\cellcolor{gray!25} & 98.16\cellcolor{gray!25} & 0.4752\cellcolor{gray!25} & 96.55\cellcolor{gray!25} \\ \cline{2-8} 
          & Utt. & 99.58\cellcolor{gray!25} & 99.17\cellcolor{gray!25} & 99.37\cellcolor{gray!25} & 99.89\cellcolor{gray!25} & 0.8160\cellcolor{gray!25} & 97.39\cellcolor{gray!25} \\\hline

    \end{tabular}} \\
    \vspace{3mm}
    
    (b) LlamaPartialSpoof (out-of-domain) \\
    \scalebox{0.92}{
    \begin{tabular}{ll|rrr|rrr}
         \hline \hline
         \multirow{2}{*}{Model} & \multirow{2}{*}{Reso.} &  \multicolumn{3}{c|}{Threshold=0.5000} & \multicolumn{3}{c}{Recall=90\%} \\
          & & Pre. & Rec. & F1 & Pre. & Thres. & F1 \\ \hline
          \multirow{7}{*}{MRM} & 0.02 s & 71.76 & 44.42 & 54.87 & 67.98 & 0.0072 & 77.66 \\ 
          & 0.04 s & 72.71 & 43.52 & 54.45 & 68.46 & 0.0040 & 77.89 \\ 
          & 0.08 s & 74.31 & 44.84 & 55.93 & 69.46 & -0.0132 & 78.49 \\ 
          & 0.16 s & 76.98 & 48.24 & 59.31 & 71.48 & -0.0012 & 79.88 \\ 
          & 0.32 s & 80.54 & 54.23 & 64.82 & 74.14 & 0.0048 & 81.34 \\ 
          & 0.64 s & 84.28 & 64.20 & 72.88 & 78.12 & 0.0116 & 83.70 \\ \cline{2-8}
          & Utt. & 88.35 & 98.24 & 93.03 & 92.01 & 0.8652 & 91.00 \\\hline

          \multirow{7}{*}{CFPRF} & 0.02 s & 72.47 & 68.54 & 70.45 & 69.68 & 0.0092 & 78.59 \\ 
          & 0.04 s & 72.92\cellcolor{gray!25} & 69.64\cellcolor{gray!25} & 71.24\cellcolor{gray!25} & 70.17\cellcolor{gray!25} & 0.0144\cellcolor{gray!25} & 78.86\cellcolor{gray!25} \\
          & 0.08 s & 73.69\cellcolor{gray!25} & 71.58\cellcolor{gray!25} & 72.62\cellcolor{gray!25} & 71.03\cellcolor{gray!25} & 0.0204\cellcolor{gray!25} & 79.45\cellcolor{gray!25} \\ 
          & 0.16 s & 74.95\cellcolor{gray!25} & 74.77\cellcolor{gray!25} & 74.86\cellcolor{gray!25} & 72.50\cellcolor{gray!25} & 0.0304\cellcolor{gray!25} & 80.31\cellcolor{gray!25} \\ 
          & 0.32 s & 76.81\cellcolor{gray!25} & 79.49\cellcolor{gray!25} & 78.13\cellcolor{gray!25} & 74.82\cellcolor{gray!25} & 0.0544\cellcolor{gray!25} & 81.71\cellcolor{gray!25} \\ 
          & 0.64 s & 79.44\cellcolor{gray!25} & 85.49\cellcolor{gray!25} & 82.35\cellcolor{gray!25} & 78.28\cellcolor{gray!25} & 0.1896\cellcolor{gray!25} & 83.73\cellcolor{gray!25} \\ \cline{2-8} 
          & Utt. & 86.30\cellcolor{gray!25} & 99.65\cellcolor{gray!25} & 92.50\cellcolor{gray!25} & 89.92\cellcolor{gray!25} & 0.9852\cellcolor{gray!25} & 90.00\cellcolor{gray!25} \\\hline

          \multirow{7}{*}{reCFPRF} & 0.02 s & 75.75 & 50.68 & 60.73 & 69.63 & -0.0348 & 78.56 \\ 
          & 0.04 s & 76.33\cellcolor{gray!25} & 52.31\cellcolor{gray!25} & 62.08\cellcolor{gray!25} & 70.14\cellcolor{gray!25} & -0.0288\cellcolor{gray!25} & 78.86\cellcolor{gray!25} \\
          & 0.08 s & 77.30\cellcolor{gray!25} & 55.00\cellcolor{gray!25} & 64.27\cellcolor{gray!25} & 71.06\cellcolor{gray!25} & -0.0216\cellcolor{gray!25} & 79.44\cellcolor{gray!25} \\ 
          & 0.16 s & 78.68\cellcolor{gray!25} & 59.51\cellcolor{gray!25} & 67.77\cellcolor{gray!25} & 72.66\cellcolor{gray!25} & -0.0124\cellcolor{gray!25} & 80.41\cellcolor{gray!25} \\ 
          & 0.32 s & 80.45\cellcolor{gray!25} & 66.56\cellcolor{gray!25} & 72.85\cellcolor{gray!25} & 75.25\cellcolor{gray!25} & 0.0012\cellcolor{gray!25} & 81.99\cellcolor{gray!25} \\ 
          & 0.64 s & 82.59\cellcolor{gray!25} & 75.92\cellcolor{gray!25} & 79.11\cellcolor{gray!25} & 78.93\cellcolor{gray!25} & 0.0404\cellcolor{gray!25} & 84.11\cellcolor{gray!25} \\ \cline{2-8} 
          & Utt. & 86.51\cellcolor{gray!25} & 98.67\cellcolor{gray!25} & 92.19\cellcolor{gray!25} & 91.08\cellcolor{gray!25} & 0.9464\cellcolor{gray!25} & 90.54\cellcolor{gray!25} \\\hline
    \end{tabular}} \\
    \vspace{3mm}

    (c) Half-Truth (out-of-domain) \\
    \scalebox{0.92}{
    \begin{tabular}{ll|rrr|rrr}
         \hline \hline
         \multirow{2}{*}{Model} & \multirow{2}{*}{Reso.} &  \multicolumn{3}{c|}{Threshold=0.5000} & \multicolumn{3}{c}{Recall=90\%} \\
          & & Pre. & Rec. & F1 & Pre. & Thres. & F1 \\ \hline
          \multirow{7}{*}{MRM} & 0.02 s & 18.73 & 96.67 & 31.38 & 19.38 & 0.9664 & 31.90 \\ 
          & 0.04 s & 19.22 & 96.14 & 32.04 & 19.79 & 0.9440 & 32.44 \\ 
          & 0.08 s & 19.96 & 96.40 & 33.07 & 20.62 & 0.9644 & 33.56 \\ 
          & 0.16 s & 21.52 & 97.28 & 35.24 & 22.42 & 0.9604 & 35.91 \\ 
          & 0.32 s & 24.56 & 98.68 & 39.33 & 25.69 & 0.9720 & 39.98 \\ 
          & 0.64 s & 29.83 & 99.76 & 45.93 & 31.04 & 0.9656 & 46.17 \\\hline

          \multirow{7}{*}{CFPRF} & 0.02 s & 24.55 & 99.28 & 39.36 & 32.51 & 0.9444 & 47.78 \\ 
          & 0.04 s & 24.83\cellcolor{gray!25} & 99.40\cellcolor{gray!25} & 39.74\cellcolor{gray!25} & 33.27\cellcolor{gray!25} & 0.9536\cellcolor{gray!25} & 48.60\cellcolor{gray!25} \\
          & 0.08 s & 25.28\cellcolor{gray!25} & 99.57\cellcolor{gray!25} & 40.32\cellcolor{gray!25} & 34.58\cellcolor{gray!25} & 0.9648\cellcolor{gray!25} & 49.96\cellcolor{gray!25} \\ 
          & 0.16 s & 26.41\cellcolor{gray!25} & 99.72\cellcolor{gray!25} & 41.76\cellcolor{gray!25} & 36.48\cellcolor{gray!25} & 0.9756\cellcolor{gray!25} & 51.93\cellcolor{gray!25} \\ 
          & 0.32 s & 28.83\cellcolor{gray!25} & 99.92\cellcolor{gray!25} & 44.75\cellcolor{gray!25} & 39.54\cellcolor{gray!25} & 0.9852\cellcolor{gray!25} & 54.98\cellcolor{gray!25} \\ 
          & 0.64 s & 33.33\cellcolor{gray!25} & 100.00\cellcolor{gray!25} & 50.00\cellcolor{gray!25} & 44.23\cellcolor{gray!25} & 0.9928\cellcolor{gray!25} & 59.40\cellcolor{gray!25} \\\hline

          \multirow{7}{*}{reCFPRF} & 0.02 s & 35.27 & 95.57 & 51.52 & 48.87 & 0.9784 & 63.34 \\ 
          & 0.04 s & 35.37\cellcolor{gray!25} & 95.89\cellcolor{gray!25} & 51.68\cellcolor{gray!25} & 50.42\cellcolor{gray!25} & 1.0012\cellcolor{gray!25} & 64.63\cellcolor{gray!25} \\
          & 0.08 s & 35.59\cellcolor{gray!25} & 96.52\cellcolor{gray!25} & 52.00\cellcolor{gray!25} & 52.98\cellcolor{gray!25} & 1.0296\cellcolor{gray!25} & 66.70\cellcolor{gray!25} \\ 
          & 0.16 s & 36.48\cellcolor{gray!25} & 97.52\cellcolor{gray!25} & 53.10\cellcolor{gray!25} & 56.23\cellcolor{gray!25} & 1.0536\cellcolor{gray!25} & 69.23\cellcolor{gray!25} \\ 
          & 0.32 s & 38.36\cellcolor{gray!25} & 98.67\cellcolor{gray!25} & 55.25\cellcolor{gray!25} & 60.81\cellcolor{gray!25} & 1.0744\cellcolor{gray!25} & 72.58\cellcolor{gray!25} \\ 
          & 0.64 s & 41.54\cellcolor{gray!25} & 99.32\cellcolor{gray!25} & 58.58\cellcolor{gray!25} & 65.43\cellcolor{gray!25} & 1.0904\cellcolor{gray!25} & 75.82\cellcolor{gray!25} \\ \hline
    \end{tabular}} \\
    \vspace{-6mm}
    \label{tab:llamapartialspoof}
\end{table}

\section{Experiments}
\subsection{Detecting fake segment at different resolutions}

In the first experiment, we explored the trade-off between precisely localizing fake segments and simply detecting their presence. In other words, we evaluated the performance across different segment durations. Detecting shorter segments is more challenging and requires a more precise system; however, in real-world applications, such precision may be less valuable than overall robustness.
We selected the MRM model for our experiments because it produces predictions at multiple resolutions. Although the CFPRF model achieved a lower EER at 20-ms resolution \cite{wu2024coarse}, it does not natively output scores for longer segments. To obtain scores at coarser resolutions, we downsampled the CFPRF score sequence by a factor of n and aggregated adjacent segments using the maximum value, a simple heuristic that has been shown to be effective \cite{zhang23v_interspeech}.

Table \ref{tab:eer} shows the segment-based EER at various resolutions on several in-domain and out-of-domain evaluation sets. It includes the MRM model, the original CFPRF model provided by the authors, and our reproduced version (reCFPRF), all trained on the training set of the PartialSpoof dataset. As expected, performance generally improves with larger segment sizes, suggesting that increasing segment length is a valid trade-off for enhancing robustness.
Interestingly, our reproduced model, reCFPRF, achieved higher EER on in-domain data but lower EER on both out-of-domain sets\footnote{Half-Truth does not have utterance-based EER since its test set does not contains bona fide samples} compared to the original CFPRF. This suggests that the original model may have been overfitted to its training data.

\begin{table}[t]
    \caption{Effects of waveform augmentation on generalization ability of the localization model at the 20-ms resolution}
    \centering

    (a) PartialSpoof (in-domain) \\
    
    \scalebox{1.0}{
    \begin{tabular}{l|r|rrrrr}
         \hline \hline
         \multirow{2}{*}{Model} & \multirow{2}{*}{EER} & \multicolumn{5}{c}{Threshold at EER}  \\
          &  & Thres. & Acc. & Pre. & Rec. & F1 \\ \hline
         MRM & 13.72 & 0.0124 & 86.24 & 80.20 & 86.46 & 83.21 \\
         MRM+rb & 15.27 & \textbf{0.0924} & 84.73 & 78.34 & 84.70 & 81.40 \\ 
         MRM+ms & \textbf{12.86} & 0.0084 & \textbf{87.11} & \textbf{81.39} & \textbf{87.26} & \textbf{84.22} \\ \hline
         CFPRF (*) & 7.61 & 0.0880 & 92.39 & 88.77 & 92.39 & 90.54 \\ \hline
         reCFPRF & 9.84 & 0.0504 & 90.15 & 85.62 & 90.16 & 87.83 \\
         reCFPRF+rb & \textbf{9.52} & \textbf{0.1892} & \textbf{90.48} & \textbf{86.09} & \textbf{90.48} & \textbf{88.23} \\
         reCFPRF+ms & 11.92 & 0.0800 & 88.07 & 82.76 & 88.11 & 85.35 \\ \hline
    \end{tabular}} \\
    \vspace{3mm}
    
    (b) LlamaPartialSpoof (out-of-domain) \\
    \scalebox{1.0}{
    \begin{tabular}{l|r|rrrrr}
         \hline \hline
         \multirow{2}{*}{Model} & \multirow{2}{*}{EER} & \multicolumn{5}{c}{Threshold at EER}  \\
          &  & Thres. & Acc. & Pre. & Rec. & F1 \\ \hline
         MRM & 46.29 & 0.0548 & 53.71 & 71.36 & 53.71 & 61.29 \\
         MRM+rb & 47.88 & \textbf{0.0764} & 52.12 & 70.04 & 52.11 & 59.76 \\ 
         MRM+ms & \textbf{44.70} & 0.0196 & \textbf{55.33} & \textbf{72.64} & \textbf{55.40} & \textbf{62.86} \\ \hline
         CFPRF (*) & 43.25 & 0.8828 & 56.75 & 73.81 & 56.76 & 64.17 \\ \hline
         reCFPRF & \textbf{41.72} & \textbf{0.2444} & \textbf{58.28} & \textbf{75.00} & \textbf{58.28} & \textbf{65.59} \\
         reCFPRF+rb & 42.59 & 0.0844 & 56.41 & 73.55 & 56.39 & 63.84 \\
         reCFPRF+ms & 42.23 & 0.1052 & 57.77 & 74.60 & 57.78 & 65.12 \\ \hline
    \end{tabular}} \\
    \vspace{3mm}

    (c) Half-Truth (out-of-domain) \\
    \scalebox{1.0}{
    \begin{tabular}{l|r|rrrrr}
         \hline \hline
         \multirow{2}{*}{Model} & \multirow{2}{*}{EER} & \multicolumn{5}{c}{Threshold at EER}  \\
          &  & Thres. & Acc. & Pre. & Rec. & F1 \\ \hline
         MRM & 46.48 & 0.9908 & 53.44 & 20.65 & 53.64 & 29.82 \\
         MRM+rb & \textbf{38.40} & 0.9728 & 61.57 & 26.61 & \textbf{61.65} & \textbf{37.17} \\ 
         MRM+ms & 38.51 & \textbf{0.9956} & \textbf{62.32} & \textbf{26.79} & 60.19 & 37.07 \\ \hline
         CFPRF (*) & 27.59 & 0.9704 & 72.34 & 37.19 & 72.52 & 49.16 \\ \hline
         reCFPRF & 14.98 & \textbf{1.0420} & 84.97 & 56.10 & 85.12 & 67.63 \\
         reCFPRF+rb & 13.44 & 0.9128 & 86.56 & 59.29 & 86.55 & 70.37 \\
         reCFPRF+ms & \textbf{12.00} & 0.8280 & \textbf{88.01} & \textbf{62.42} & \textbf{87.99} & \textbf{73.03} \\ \hline
    \end{tabular}} \\
    \vspace{1mm}
    \label{tab:augmentation}
    \footnotesize{(*) Since it is a training from scratch experiments, the pretrained CFPRF did not include waveform augmentation results.}
    \vspace{-5mm}
\end{table}

\subsection{Threshold, Recall, Precision, and F1}
The problem with using EER to evaluate performance across multiple datasets is that each result corresponds to different decision thresholds. In other words, models that achieve strong EER scores across different evaluation sets may still perform poorly in deployment which requires a single threshold.
Since all of our models output score value approximately between 0 and 1 (uncapped, as logits were used), we can use a neutral threshold of 0.5 for evaluation. Table \ref{tab:eer} reports the Precision, Recall, and F1 of the baseline models at different resolutions. Interestingly, while the PartialSpoof and LlamaPartialSpoof datasets exhibit higher precision than recall, the Half-Truth test set shows the opposite trend.

Moreover, since deepfake detection typically operates under the assumption that a false alarm is far less costly than a missed detection \cite{wang2024asvspoof}, we evaluate models at a high recall level, 95\% for in-domain and 90\% for out-of-domain data, and compare their precision at these points.
For PartialSpoof evaluated at 20-ms resolution and a 95\% recall level, MRM achieves 53.95\% precision, while CFPRF and reCFPRF reach 81.35\% and 64.28\%, respectively. However, on LlamaPartialSpoof at 90\% recall, the differences in precision among the models are smaller. On the Half-Truth dataset, reCFPRF performs better than CFPRF under the same recall condition. 
It is also important to note that the thresholds required to achieve a specific recall value vary across datasets. In real-world applications, one may use a neutral threshold (e.g., 0.5) or select a threshold based on a specific dataset. However, in the latter case, an optimal threshold for in-domain evaluation may not generalize well to out-of-domain scenarios, and vice versa. We will discuss this issue further in the next section.

\subsection{EER Threshold and Effect of Waveform Augmentation}

Since both tested models take raw waveforms as input, we investigated whether waveform augmentation methods effective in detection tasks can also benefit localization systems. Specifically, we applied the third algorithm of RawBoost \cite{tak2022rawboost} and the band-shaped MaskedSpec \cite{rimon2025unmasking} as augmentation methods, as these were reported by the authors to be the most effective in their respective works.
Table \ref{tab:augmentation} presents the EER and other evaluation results, calculated at the EER threshold, for models trained with additional waveform augmentation. Interestingly, MaskedSpec (+ms) enhances the performance of the MRM model across both in-domain and out-of-domain scenarios, while RawBoost (+rb) shows inconsistent effectiveness. For CFPRF, the effects of augmentation are inconsistent across datasets. Overall, we conclude that while waveform augmentation can enhance in-domain performance, its effectiveness in out-of-domain settings is not guaranteed.

\begin{figure*}[t]
    \begin{subfigure}[b]{0.32\textwidth}
         \centering
         \includegraphics[width=\textwidth]{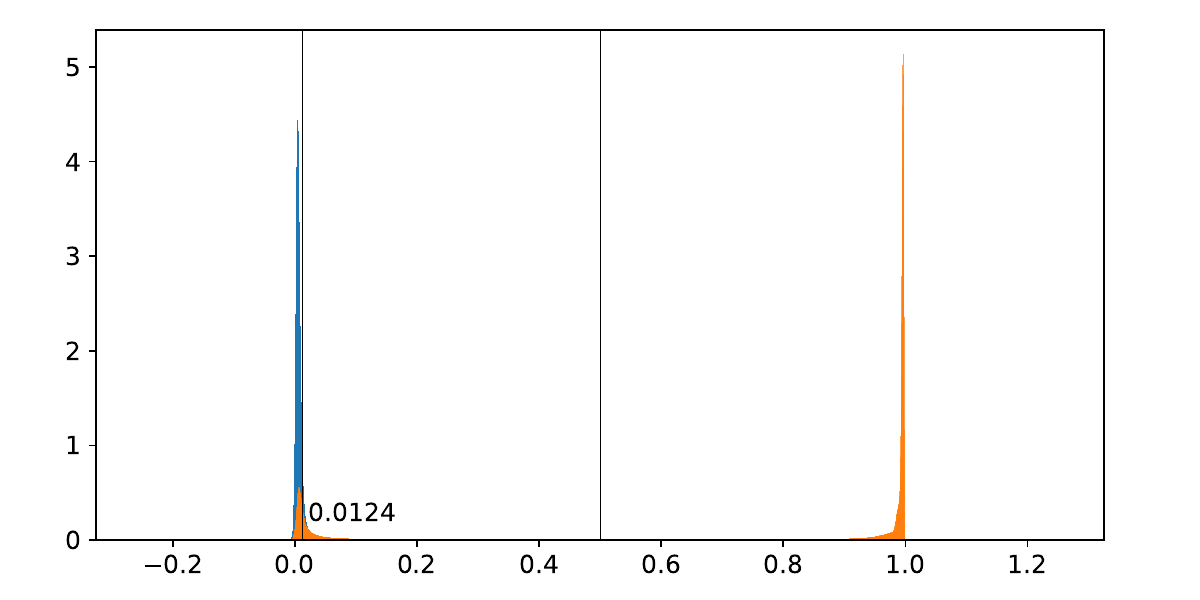} \\
         \includegraphics[width=\textwidth]{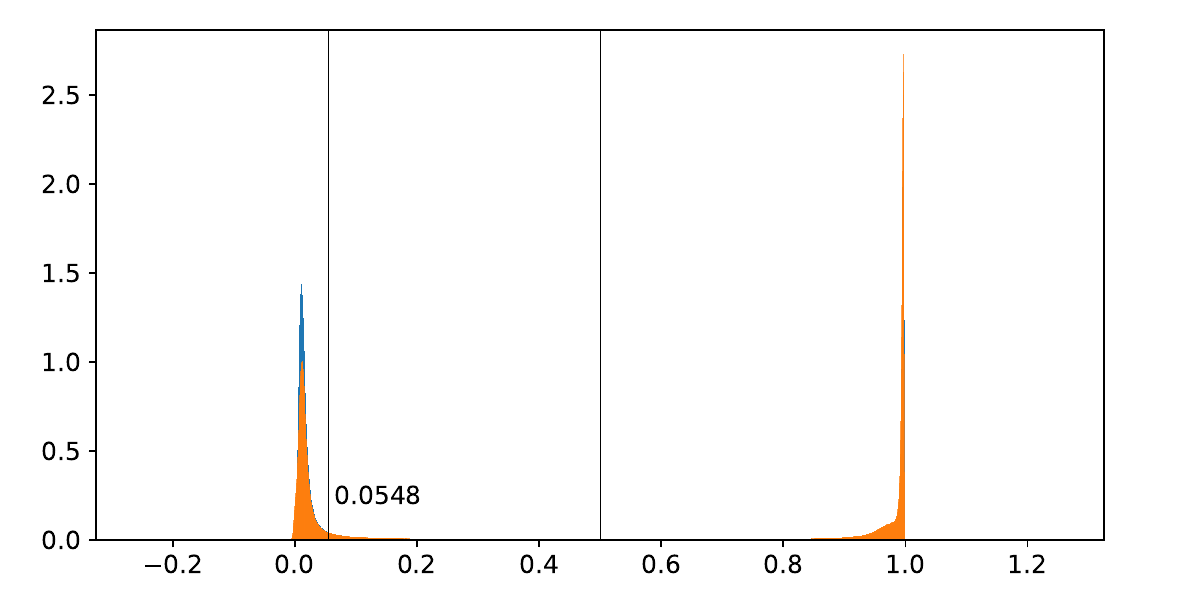}\\
         \includegraphics[width=\textwidth]{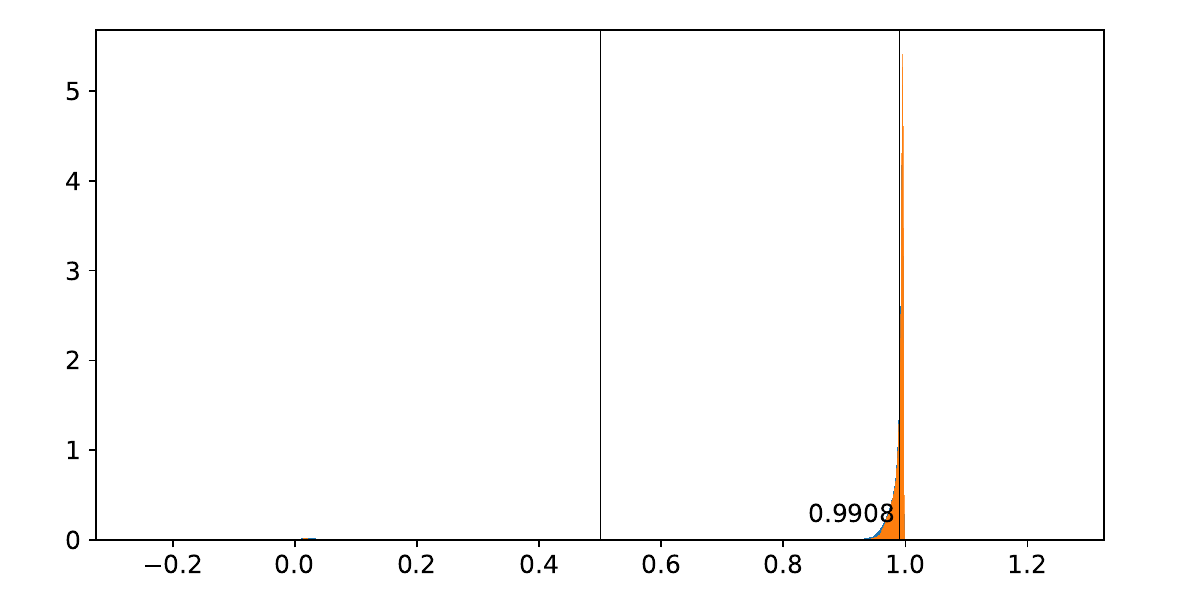}\\
         \vspace{-2mm}
         \caption{MRM}
     \end{subfigure}
     \hfill
    \begin{subfigure}[b]{0.32\textwidth}
         \centering
         \includegraphics[width=\textwidth]{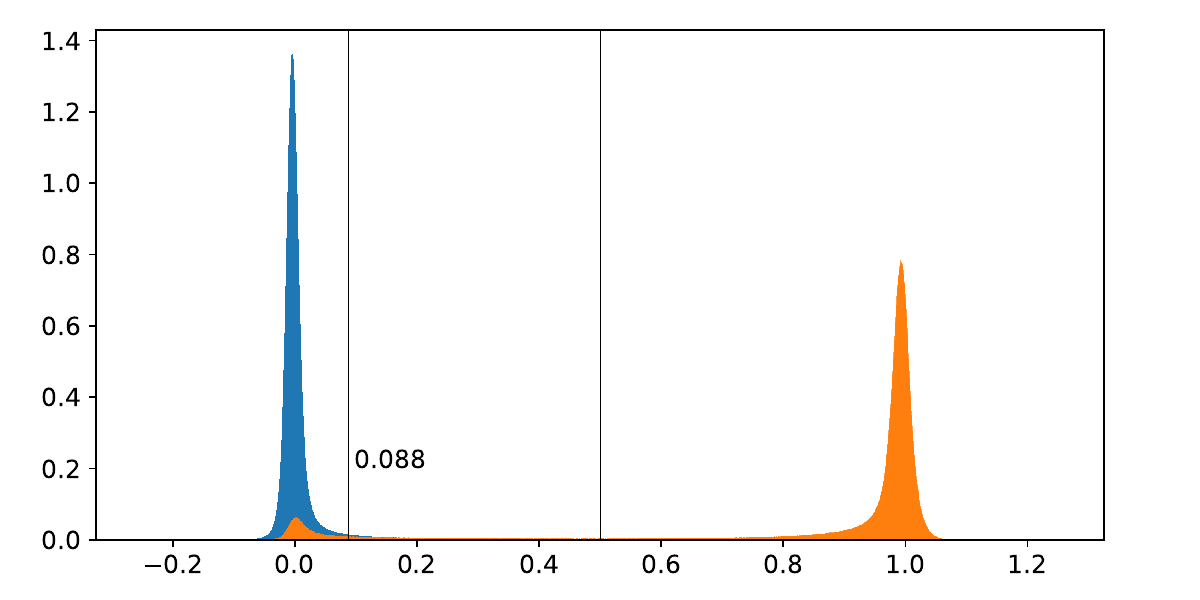} \\
         \includegraphics[width=\textwidth]{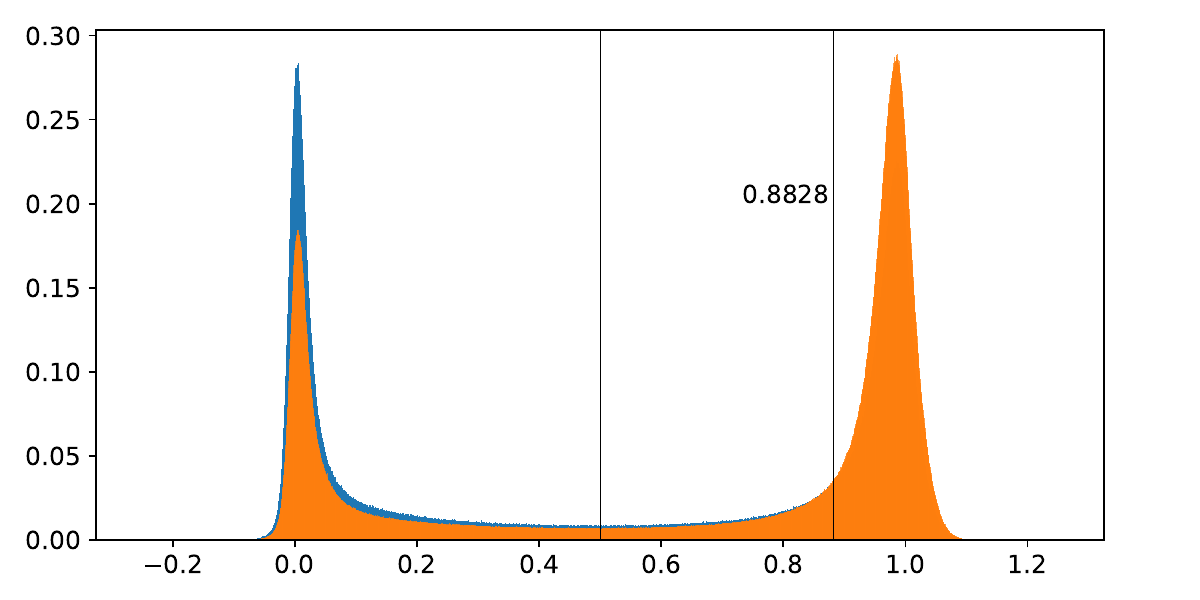}\\
         \includegraphics[width=\textwidth]{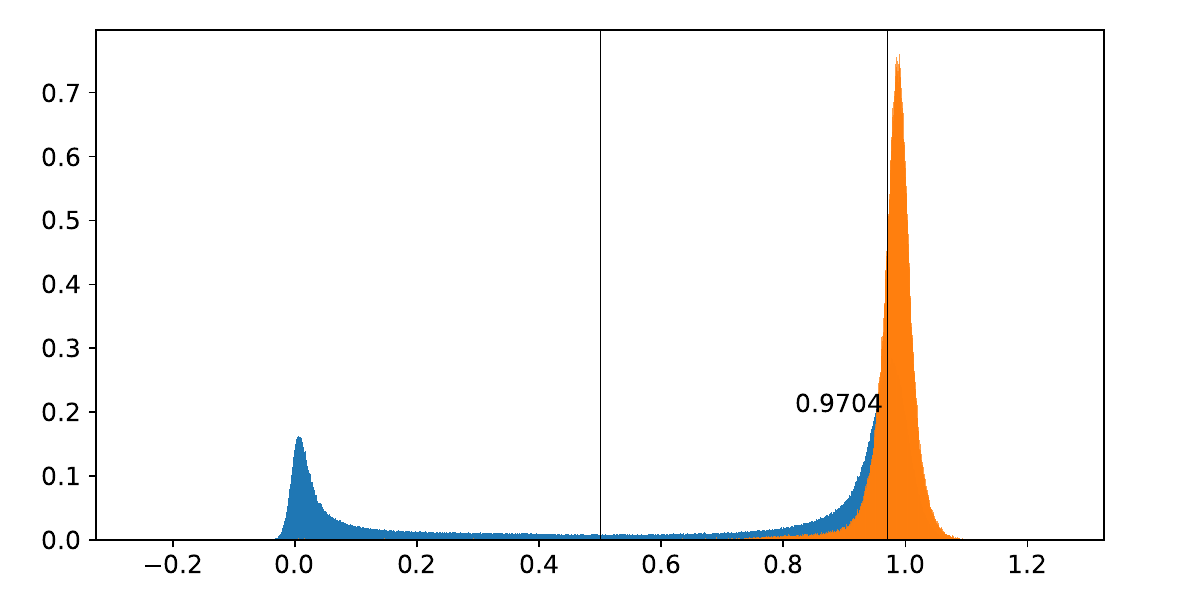}\\
         \vspace{-2mm}
         \caption{CFPRF}
     \end{subfigure}
     \hfill
    \begin{subfigure}[b]{0.32\textwidth}
         \centering
         \includegraphics[width=\textwidth]{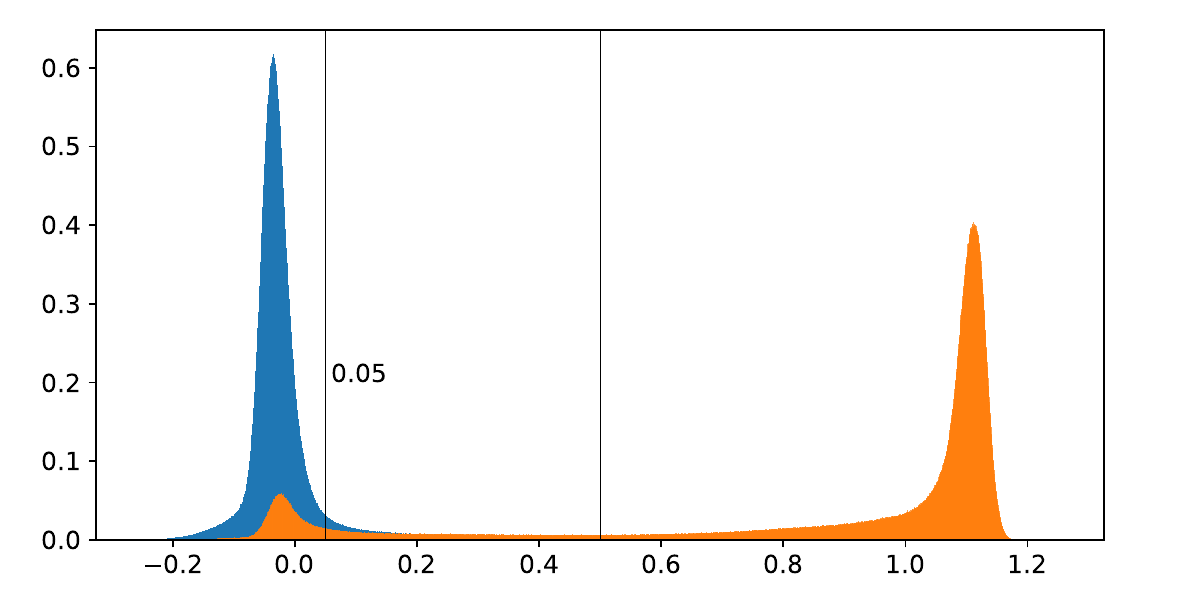} \\
         \includegraphics[width=\textwidth]{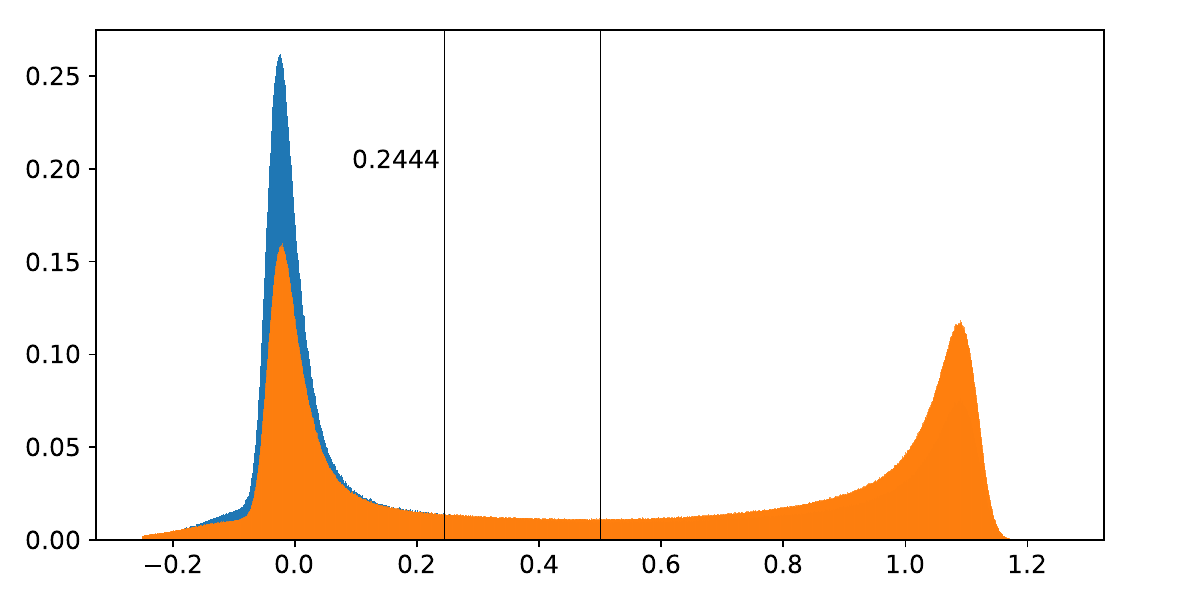}\\
         \includegraphics[width=\textwidth]{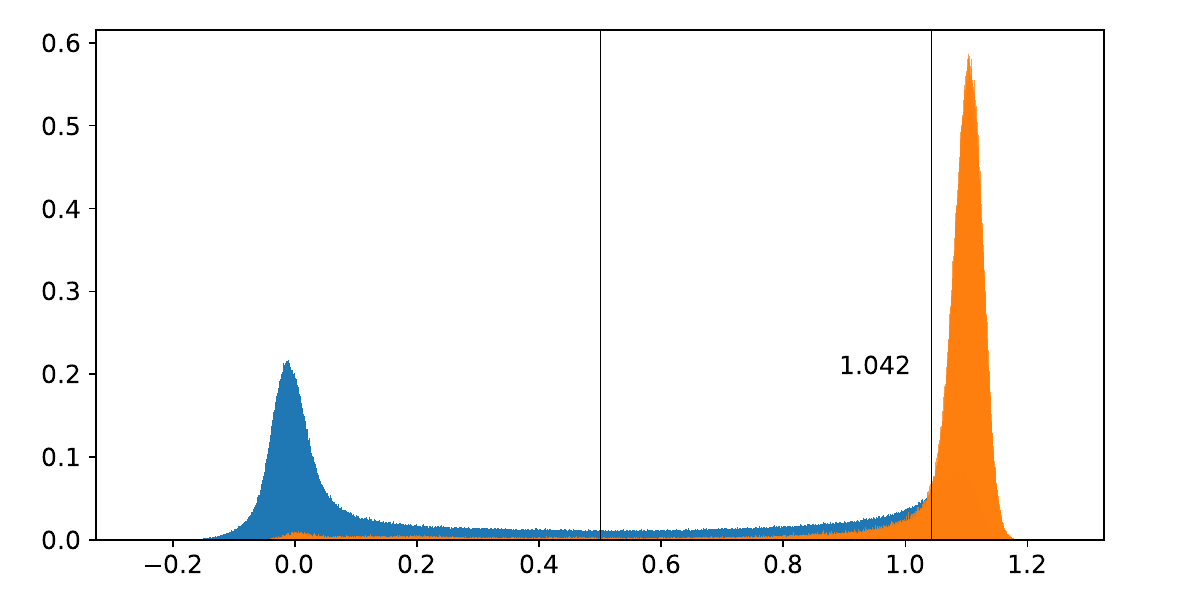}\\
         \vspace{-2mm}
         \caption{reCFPRF}
     \end{subfigure}
        \caption{Distributions of 20-ms segment score produced by the detections models. Blues are bona fide segments while oranges are fakes. Higher scores mean they are more likely to be fake. Upper row is PartialSpoof (in-domain), middle row is LlamaPartialSpoof (out-of-domain) and lower row is Half-Truth (out-of-domain). Vertical lines indicate either 0.5 threshold or EER threshold.}
        \label{fig:distribution}
  \centering
  \vspace{-3mm}
\end{figure*}

Figure \ref{fig:distribution} shows the score distributions of the three baseline models along with their EER thresholds. Among them, CFPRF stands out, as the large gap between its in-domain and out-of-domain EER thresholds suggests that it is highly overfitted to in-domain data. In practice, using the in-domain EER threshold in a production setting may result in suboptimal performance in real-world scenarios.
Moreover, we observe that all models misclassified both unseen bona fide and synthetic samples. These findings highlight a key limitation of machine learning-based approaches to fake speech detection: their strong dependence on the quality and diversity of the training data.

\subsection{Tuning with emerging data}

\begin{table}[t]
    \caption{Evaluation results of the models tuned on LlamaPartialSpoof train subset. Gray cells indicate results that is better than the original checkpoints, while bold values indicates the best within the same system type.}
    \label{tab:tuning}
    \centering

    (a) EER \\
    
    \scalebox{1.0}{
    \begin{tabular}{l|r|r|rr|r}
         \hline \hline
         \multirow{2}{*}{Model} & \multicolumn{1}{c|}{PS} & \multicolumn{3}{c|}{LPS (test)} & \multicolumn{1}{c}{HAD} \\ \cline{2-6}
          & eval & both & full & partial & test \\ \hline
         MRM & 13.72 & 44.54 & 40.98 & 32.76 & 46.48 \\ \cline{2-6}
         +lps-train-bonafide & 17.04 & 48.40 & 48.25 & 34.34 & 30.73\cellcolor{gray!25} \\ 
         +lps-train-full & \textbf{8.22}\cellcolor{gray!25} & 46.55 & 43.49 & 46.10 & 54.13 \\ 
         +lps-train-partial & 9.30\cellcolor{gray!25} & 11.39\cellcolor{gray!25} & 10.70\cellcolor{gray!25} & \textbf{9.75}\cellcolor{gray!25} & 9.62\cellcolor{gray!25} \\ 
         +lps-train-all & 9.03\cellcolor{gray!25} & \textbf{8.20}\cellcolor{gray!25} & \textbf{6.04}\cellcolor{gray!25} & 10.31\cellcolor{gray!25} & \textbf{7.55}\cellcolor{gray!25} \\ \hline
         CFPRF & \textbf{7.61} & 41.47 & 36.22 & 35.01 & 27.59 \\ \cline{2-6}
         +lps-train-bonafide & 12.14 & 47.29 & 45.98 & 35.77 & 9.53\cellcolor{gray!25} \\ 
         +lps-train-full & 7.83 & 42.66 & 41.01 & 47.40 & 44.77 \\ 
         +lps-train-partial & 8.11 & 7.62\cellcolor{gray!25} & 7.19\cellcolor{gray!25} & \textbf{6.66}\cellcolor{gray!25} & 4.74\cellcolor{gray!25} \\ 
         +lps-train-all & 7.62 & \textbf{5.96}\cellcolor{gray!25} & \textbf{2.89}\cellcolor{gray!25} & 8.72\cellcolor{gray!25} & \textbf{3.52}\cellcolor{gray!25}  \\ \hline
    \end{tabular}} \\
    \vspace{3mm}

    (b) Accuracy (Threshold=0.5) \\
    
    \scalebox{1.0}{
    \begin{tabular}{l|r|r|rr|r}
         \hline \hline
         \multirow{2}{*}{Model} & \multicolumn{1}{c|}{PS} & \multicolumn{3}{c|}{LPS (test)} & \multicolumn{1}{c}{HAD} \\ \cline{2-6}
          & eval & both & full & partial & test \\ \hline
         MRM & 87.34 & 52.04 & 47.73 & 66.98 & 22.02 \\ \cline{2-6}
         +lps-train-bonafide & 83.96 & 38.69 & 31.07 & 65.45 & 51.52\cellcolor{gray!25}\\ 
         +lps-train-full & \textbf{92.34}\cellcolor{gray!25} & 65.06\cellcolor{gray!25} & 71.25\cellcolor{gray!25} & 38.96 & 18.46 \\ 
         +lps-train-partial & 90.36\cellcolor{gray!25} & 89.17\cellcolor{gray!25} & 89.27\cellcolor{gray!25} & \textbf{89.12}\cellcolor{gray!25} & 74.39\cellcolor{gray!25}  \\ 
         +lps-train-all & 90.45\cellcolor{gray!25} & \textbf{92.37}\cellcolor{gray!25} & \textbf{94.62}\cellcolor{gray!25} & 88.99\cellcolor{gray!25} & \textbf{84.49}\cellcolor{gray!25}\\ \hline
         CFPRF & 93.00 & 61.19 & 62.86 & 61.27 & 43.59 \\ \cline{2-6}
         +lps-train-bonafide & 90.29 & 41.26 & 34.63 & 66.00\cellcolor{gray!25} & 90.69\cellcolor{gray!25} \\ 
         +lps-train-full & 93.03\cellcolor{gray!25} & 64.98\cellcolor{gray!25} & 71.09\cellcolor{gray!25} & 38.68 & 18.60 \\ 
         +lps-train-partial & 93.33\cellcolor{gray!25} & 92.85\cellcolor{gray!25} & 93.20\cellcolor{gray!25} & \textbf{92.07}\cellcolor{gray!25} & 92.21\cellcolor{gray!25} \\ 
         +lps-train-all & \textbf{93.59}\cellcolor{gray!25} & \textbf{93.43}\cellcolor{gray!25} & \textbf{95.13}\cellcolor{gray!25} & 88.71\cellcolor{gray!25} & \textbf{94.86}\cellcolor{gray!25} \\ \hline
    \end{tabular}} \\
    \vspace{-3mm}
\end{table}

Previous sections have shown that current SSL-based localization methods perform well on in-domain data but generalize poorly to out-of-domain scenarios. While it is often assumed that training on large-scale and diverse datasets can resolve generalization issues, we test this hypothesis by simulating a scenario in which new data becomes available for tuning.
Specifically, we selected approximately 20 speakers from the LlamaPartialSpoof (LPS) dataset as the new train subset for fine-tuning, with the remaining speakers used as the new test set, as shown in Table \ref{tab:data}.
Starting from the checkpoints of the two baseline systems trained on PartialSpoof (PS), we fine-tuned the models using a combination of the PS train data and either part or all of the LPS train subset.

Table \ref{tab:tuning} presents the EER and the accuracy of the baseline and fine-tuned models on the PS eval set, the new LPS test subset (including separate evaluations on fully and partially fake utterances), and the Half-Truth Audio Deepfake (HAD) test set.
Interestingly, adding bona fide or fully synthetic utterances to the training data often worsens performance across most metrics, contradicting the common belief. It is worth noting that adding bona fide utterances does improve the performance on the HAD test set, while adding fully synthetic utterances improves the performance of the MRM model on the PS eval set, even though they are largely unrelated.
More importantly, adding partial spoof utterances from the LPS train set leads to significant improvements across most test conditions, suggesting that these models learn to distinguish between real and fake segments based on temporal inconsistencies within a single utterance. As expected, models fine-tuned with all three types of samples produce the most consistent results overall.
This suggests that simply increasing the amount of training data does not guarantee improved performance. A thoughtful data creation and selection strategy may be necessary to develop reliable localization systems.

\section{Conclusions}

In this work, we revisited the evaluation setup for partially fake speech localization and identified key gaps between academic benchmarks and real-world deployment, both in terms of metrics and dataset diversity. We argued that metrics like EER, while useful for benchmarking, are insufficient for interpreting real-world system behavior. To address this, we reframe the localization task as a sequential anomaly detection problem and advocated for the use of more interpretable, threshold-dependent metrics.
We further showed that current SSL-based models, despite strong in-domain performance, generalize poorly to out-of-domain data. These models not only fail to detect novel synthetic samples but also misclassify unfamiliar bona fide audio, highlighting their reliance on domain-specific patterns.
For future work, we recommend always including an out-of-domain evaluation set, such as LlamaPartialSpoof, and focusing on areas such as domain adaptation \cite{li2024cross,chen2025continual} and generalizable training methods \cite{doan2024trident,lu2025leveraging}.
Such a shift is critical for advancing the field toward robust solutions that withstand increasingly sophisticated synthetic speech \cite{vyas2023audiobox} and adversarial attacks \cite{muller2025replay} in real-world conditions.

% \printbibliography

\bibliographystyle{IEEEtran}
\bibliography{mybib}

\end{document}